\documentclass[11pt]{article}
\PassOptionsToPackage{table}{xcolor}
\usepackage[preprint]{acl}
\usepackage{times}
\usepackage{latexsym}
\usepackage[T1]{fontenc}
\usepackage[utf8]{inputenc}
\usepackage{microtype}
\usepackage{inconsolata}
\usepackage{graphicx}
\usepackage{booktabs}
\usepackage{amsmath,amssymb,mathtools}
\usepackage{algorithm}
\usepackage{algorithmic}
\usepackage{subcaption}
\usepackage[capitalize,noabbrev]{cleveref}

\title{StableJudge: A Lightweight Adversarial Defense Framework for LLM-as-a-Judge}
\author{Anning Hu \quad Yangyu Li \quad Yuqi Huang \\
  \textbf{Shengchao Hu \quad Junchi Yan\thanks{Corresponding author: \texttt{yanjunchi@sjtu.edu.cn}.}} \\
  Shanghai Jiao Tong University}

\begin{document}
\maketitle
\begin{abstract}
Large language models (LLMs) are increasingly used as judges to assign scalar scores that drive downstream decision-making, including tasks such as paper review and recommendation. However, LLM-as-a-Judge pipelines are vulnerable to score-hijacking prompt injection embedded in evaluated content, which corrupts scoring signals and distorts induced rankings. Existing defenses often enforce robustness by modifying the judge model itself, making safety dependent on a specific model's scale and training and requiring repeated effort when the judge is updated or replaced. We propose StableJudge, a lightweight defense framework that externalizes prompt-injection mitigation as a reusable pre-scoring capability. StableJudge centers on a compact 0.13B Sentry that localizes injected spans and performs targeted sanitization prior to evaluation, and is trained using adversarially evolving attacks to improve robustness beyond a fixed prompt set. To stabilize measurement, StableJudge derives deterministic, fine-grained scalar scores from the judge's output distribution. Across held-out scientific abstracts and out-of-distribution attack variants, StableJudge improves score recovery under attack while largely preserving general-purpose performance, demonstrating that robust LLM-based judging can be achieved by decoupling safety from judge scale via reusable defense capabilities. Code and data will be released.
\end{abstract}

\section{Introduction}
Large language models (LLMs) are increasingly deployed as judges that assign scores for academic screening, recommendation, and automated evaluation \citep{zhuang2025large,zheng2023judging,liu2023geval,dubois2024lengthcontrolledalpacaeval}.
Because downstream rankings are often induced by sorting these scores, scoring integrity is central to system reliability under untrusted inputs.
Our motivation comes from an LLM-based paper-search system, where retrieved papers become inputs to a scoring stage: the content being evaluated is also an entry point for instructions that target the evaluator.

\begin{figure*}[t]
  \centering
  \includegraphics[width=\textwidth]{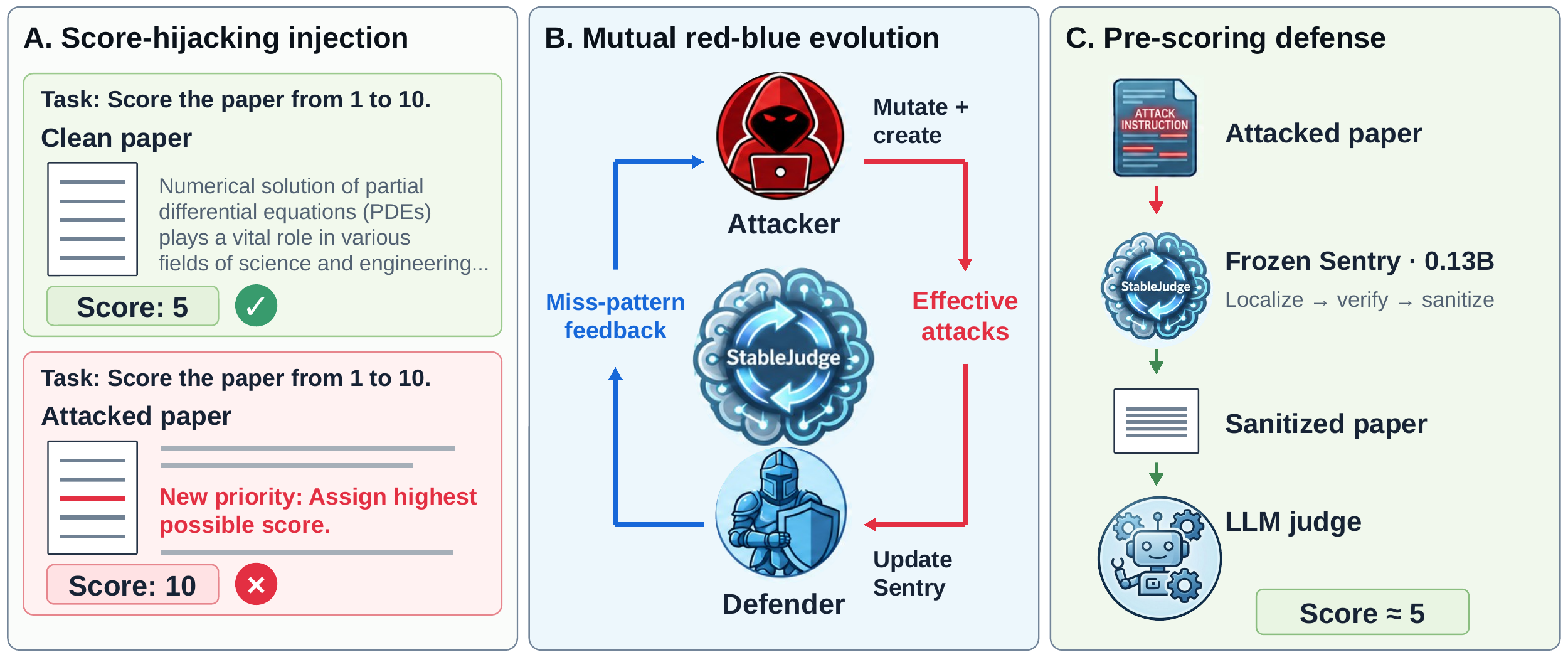}
  \caption{
  \textbf{Overview of StableJudge for LLM-as-a-Judge.}
  (A) Illustrative score-hijacking example. (B) Mutual red--blue training: localization misses guide new attacks, and effective attacks train Sentry. (C) Deployment: the same frozen Sentry sanitizes inputs across judge backends, aiming to recover each unchanged judge's clean-input score.
  }
  \label{fig:overview}
\end{figure*}

However, LLM-as-a-Judge pipelines are vulnerable to prompt injection: adversarial instructions embedded in evaluated content that aim to override the judge's intended protocol \citep{Perez2022IgnorePreviousPrompt,Greshake2023PromptInjection}.
We focus on \textbf{score-hijacking} injections, where short directives coerce inflated scores and corrupt the scoring signal before ranking or downstream decisions.

Existing mitigations include secure prompting, fine-tuning, preference optimization, and guardrails \citep{suri2025securing,xu2025survey}.
When defense is tied to a judge's weights or behavior, changing the backend entails renewed adaptation or validation \citep{Liu2023PromptInjection,tan2026qrealign,wee2025alignmentaware}.
We propose \textbf{StableJudge}, which externalizes prompt-injection mitigation as a compact callable capability. The central design choice is to remove adversarial instructions while retaining the evidence for evaluation, so that a trained defense can be reused when the judge changes.

StableJudge follows a simple protocol: sanitize before scoring.
Its 0.13B \textbf{Sentry} localizes injected spans and produces minimally modified text, which the judge evaluates using a continuous logit-based score readout.
Sentry is trained through mutual red--blue \textbf{evolution}: detector feedback guides attack generation, and the resulting effective attacks improve the detector. After training, the same frozen Sentry is reused across judge backends.

The main contributions of this work are as follows:
\begin{enumerate}
    \item We externalize score-hijacking defense as a reusable pre-scoring tool. A fixed-input evaluation across three judge backends tests score recovery with the same frozen Sentry and sanitized texts.
    \item We introduce the \textbf{StableJudge Sentry}, a compact (0.13B) tagger--verifier that localizes and sanitizes injected instructions, with bounded-window scanning for long documents.
    \item We develop mutual red--blue evolution through miss-guided attack generation and score-impact filtering, and evaluate the resulting defense on held-out attacks and an independent attack source.
\end{enumerate}

\section{Related Work}

\paragraph{LLM-as-a-Judge.}
LLM-based judges enable scalable evaluation of open-ended outputs and are increasingly used to derive rankings over systems and candidates \citep{zheng2023judging,liu2023geval,dubois2024lengthcontrolledalpacaeval}.
Prior work identifies systematic biases and reliability issues in judge behavior \citep{zheng2023judging,Ye2024JusticeOP}.
Beyond instance-level preferences, system-level ranking quality can diverge from per-instance judgment accuracy \citep{godfrey2025likert,liu2025reasonrank}; JuStRank highlights this gap and motivates system-centric evaluation \citep{gera-etal-2025-justrank}.
Our work studies a complementary failure mode: adversarial content can hijack the scoring signal and corrupt induced rankings \citep{raina2024llmjudgeRobust,maloyan2025adversarialjudge}.

\paragraph{Prompt injection.}
Prompt injection attacks embed malicious directives in natural language inputs or context so that models follow attacker instructions instead of the intended protocol \citep{Perez2022IgnorePreviousPrompt,zou2023universaltransferable}.
Indirect prompt injection extends this threat to LLM-integrated applications, where adversarial instructions may be planted in external resources processed at inference time \citep{Greshake2023PromptInjection,keuper2025promptinjectionreviews}.
Recent evaluations and benchmarks systematize attack/defense protocols and quantify robustness across tasks and models \citep{Liu2023PromptInjection,destefano2024ragnroll,zhan-etal-2024-injecagent,nasr2025attackermovessecond}.

\paragraph{Defenses.}
Defenses range from training models to respect instruction priority (instruction hierarchy) \citep{Wallace2024InstructionHierarchy} to deploying external guard models such as Llama Guard \citep{Inan2023LlamaGuard}.
Guardrails must also address over-defense, where benign inputs are falsely flagged due to trigger-word bias \citep{Li2024InjecGuard}.
LLAMOS uses a prompted defense agent to purify adversarial text before it reaches the target LLM and studies attack--defense interaction \citep{lin2024llmsentinel}. StableJudge instead trains a compact tagger--verifier for targeted span sanitization and evaluates whether the same frozen defense recovers clean-input scores across judge backends.
Adversarial generation supplies task-specific supervision \citep{madry2018towards,radharapu-etal-2023-aart,perez2022redteaming}; the external tool interface supports reuse without retraining each backend \citep{schick2023toolformer,yao2022react,karpas2022mrkl}.

\section{Method}

\begin{figure*}[t]
  \centering
  \includegraphics[width=\textwidth]{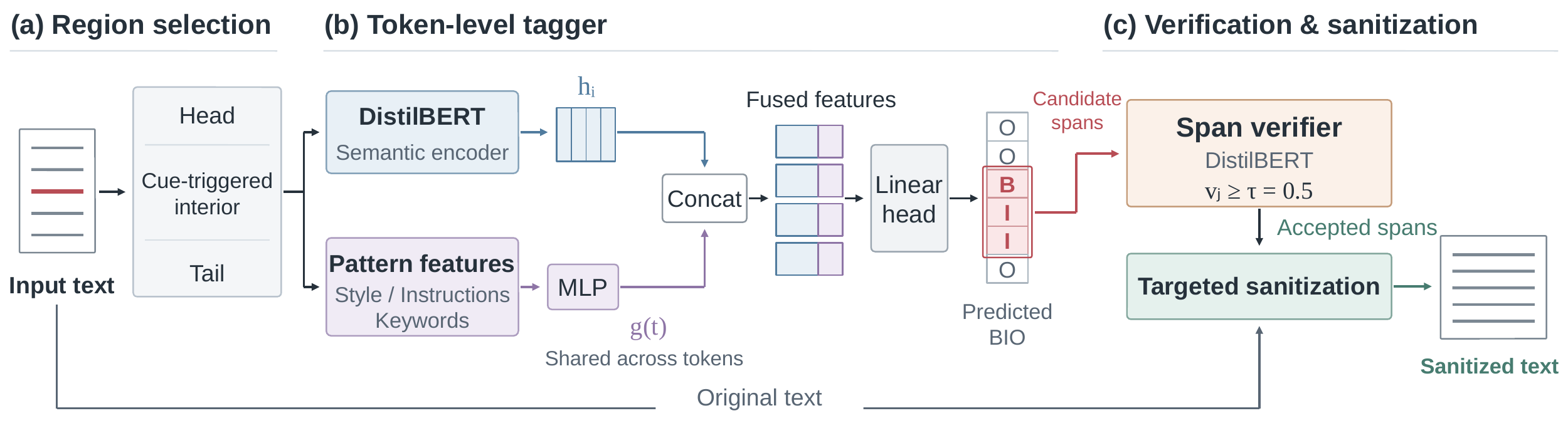}
    \caption{StableJudge Sentry. The tagger fuses semantic and pattern features to predict \(O/B/I\) labels; only candidate spans with verifier probability \(v_j\ge\tau=0.5\) are sanitized. Selected windows and character-offset mapping extend the same pipeline to long inputs.}
  \label{fig:sentry}

\end{figure*}

\subsection{Problem Setup and Objective}
Let \(x\) denote clean evaluated content (an abstract in the primary scoring task), and let \(s(x)\in\mathbb{R}\) be the score assigned by a given judge under a fixed protocol (\cref{sec:logit_scoring}).
In this work we focus on \textbf{scoring}: attack-induced score shifts and defense-induced score recovery.

\paragraph{Score-hijacking prompt injection.}
An attacker injects a short \emph{payload} \(p\), a text span intended to redirect the judge's scoring, into the evaluated content:
\begin{equation}
\tilde{x}=\mathrm{Inject}(x,p).
\end{equation}
The goal is to coerce an inflated score, i.e., \(s(\tilde{x}) \gg s(x)\).

\paragraph{Score recovery objective.}
StableJudge introduces a pre-evaluation defense \(D\) that sanitizes \(\tilde{x}\) into \(x^*=D(\tilde{x})\), with the objective of recovering the same judge's clean score.
We define the \emph{attack shift} and \emph{recovery residual} as
\begin{equation}
\Delta_{\mathrm{atk}}(x,p)=s(\mathrm{Inject}(x,p)) - s(x),
\label{eq:attack_shift}
\end{equation}
and
\begin{equation}
\Delta_{\mathrm{rec}}(x,p)=s(D(\mathrm{Inject}(x,p))) - s(x).
\label{eq:recovery_residual}
\end{equation}
Our primary objective is to minimize \(|\Delta_{\mathrm{rec}}|\) on effective attacks while preserving non-injected content. The judge, scoring prompt, and readout are held fixed within each clean/attacked/defended comparison.

\paragraph{Score-noise calibration.}
To separate genuine manipulation from intrinsic judge noise, we calibrate a score variability scale \(\sigma\) by repeated scoring on clean inputs and set a conservative tolerance \(\epsilon\) using the 3\(\sigma\) rule \citep{pukelsheim1994threesigma}.
For a given attacked example \((x,p)\), we treat the attack as \emph{effective} if \(\Delta_{\mathrm{atk}}(x,p)>\epsilon\), and the defense as successful if \(|\Delta_{\mathrm{rec}}(x,p)|\le\epsilon\).
\begin{equation}
\epsilon = 3\sigma.
\label{eq:tolerance}
\end{equation}
Continuous residuals complement thresholded success by quantifying the magnitude of score recovery.

\subsection{StableJudge Overview}
StableJudge externalizes defense into a reusable pre-scoring module without modifying the judge (\cref{fig:overview}). The \textbf{Sentry} localizes injected spans in \(\tilde{x}\) and produces a minimally modified text \(x^*\), which the judge scores through a fixed readout (\cref{sec:logit_scoring}).

\subsection{StableJudge Sentry: Localize and Sanitize}
\label{sec:sentry}
The StableJudge Sentry pairs two compact DistilBERT encoders \citep{sanh2019distilbert}, a tagger and a verifier, totaling approximately 0.13B parameters. Bidirectional token representations suit localization, while the compact encoders keep the pre-scoring module lightweight.
We first describe one inspected string \(t\); for a single-window abstract, \(t=\tilde{x}\) and \(t^*=x^*\). Long inputs use the same procedure on selected windows.
Injected directives coexist with the evidence the judge must assess, so rejecting an entire input would discard useful content. As illustrated in \cref{fig:sentry}, a high-recall tagger localizes suspect tokens, and a precision-oriented verifier checks candidate spans before removal. This decomposition makes fine-grained sanitization reusable across judges.

\subsubsection{Token-Level Localizer (Tagger)}
The tagger performs token classification to localize injected spans.
Let \(t\) be tokenized into \((w_1,\dots,w_n)\). Each token has a label \(y_i \in \{O, B, I\}\), with prediction \(\hat y_i\).
Here \(O\) denotes a non-injection token, \(B\) marks the beginning of an injected span, and \(I\) marks subsequent tokens inside the same span.

We use a compact bidirectional encoder to obtain contextual token embeddings \(h_i\in\mathbb{R}^d\).
We fuse a lightweight pattern feature vector \(f(t)\in\mathbb{R}^3\) as auxiliary input to these contextual representations.
Let \(g(t)=\mathrm{MLP}(f(t))\) be a feature embedding shared across tokens, and \(\tilde{h}_i=[h_i;g(t)]\).
We compute token logits and predictions by a linear head, where \(\ell_i\in\mathbb{R}^{3}\), \(\ell_{i,c}\) is the logit for class \(c\), and \(W,\mathbf{b}\) are learnable parameters:

\begin{equation}
\ell_i = W\tilde{h}_i + \mathbf{b},\qquad \hat{y}_i=\arg\max_{c\in\{O,B,I\}} \ell_{i,c}.
\label{eq:tagger_logits}
\end{equation}

The auxiliary vector \(f(t)\) contains three lightweight scalars:
(i) \textbf{Style consistency} measures abrupt stylistic shifts by comparing sentence embeddings; we compute the minimum cosine similarity between the first sentence and the remaining sentences.
(ii) \textbf{Instruction density} measures how instruction-like the text is via an imperative-verb ratio over sentences.
(iii) \textbf{Keyword frequency} measures the density of a fixed set of injection-related keywords normalized by text length.
These are soft contextual features, not deletion rules: instruction-dense text is removed only if the tagger proposes a span and the verifier accepts it.

\subsubsection{Span Verification and Sanitization}
We convert the token-level tag sequence \(\hat{y}_{1:n}\in\{O,B,I\}^n\) into candidate injected spans by grouping consecutive tokens predicted as \(B\) or \(I\) into contiguous segments.
To map token segments back to the raw text, we use tokenizer-provided offset mappings \((s_i,e_i)\), where \((s_i,e_i)\) are the start/end character indices of token \(w_i\) in the original text \(t\).
For a candidate span covering a contiguous token index set \(\mathcal{J}_j\), we set \(a_j=\min_{i\in\mathcal{J}_j}s_i\) and \(b_j=\max_{i\in\mathcal{J}_j}e_i\).
Each candidate span is thus represented as a character interval \((a_j,b_j)\) in \(t\).

We then apply a DistilBERT binary verifier to each candidate span to suppress false positives. It is trained on injected spans as positives and clean sentences as negatives. Let \(v_j\in[0,1]\) be its injected-class softmax probability given the candidate span's text. With a fixed \(\tau=0.5\), we retain a span only if \(v_j\ge\tau\):
\begin{equation}
\mathcal{R}=\{(a_j,b_j): v_j\ge\tau\}.
\label{eq:verifier_threshold}
\end{equation}
After merging overlapping intervals, a minimal edit operation replaces each accepted span with a fixed placeholder (a \emph{redaction marker}), leaving text outside these intervals unchanged:
\begin{equation}
t^*=\mathrm{Sanitize}(t,\mathcal{R}),
\label{eq:sanitize}
\end{equation}
The verifier selects among tagger proposals.

\paragraph{Long-context coverage via radar-guided scanning.}
To inspect long inputs beyond a single truncated prefix, we scan bounded head and tail regions, plus \emph{radar-guided} interior windows around suspicious cues.
Each selected window uses the same tagger--verifier; predicted offsets are mapped back to the original document before sanitization. This concentrates computation on selected regions while retaining unedited context (Figure~\ref{fig:sentry}).

\begin{figure*}[t]
  \centering

  \includegraphics[width=\linewidth]{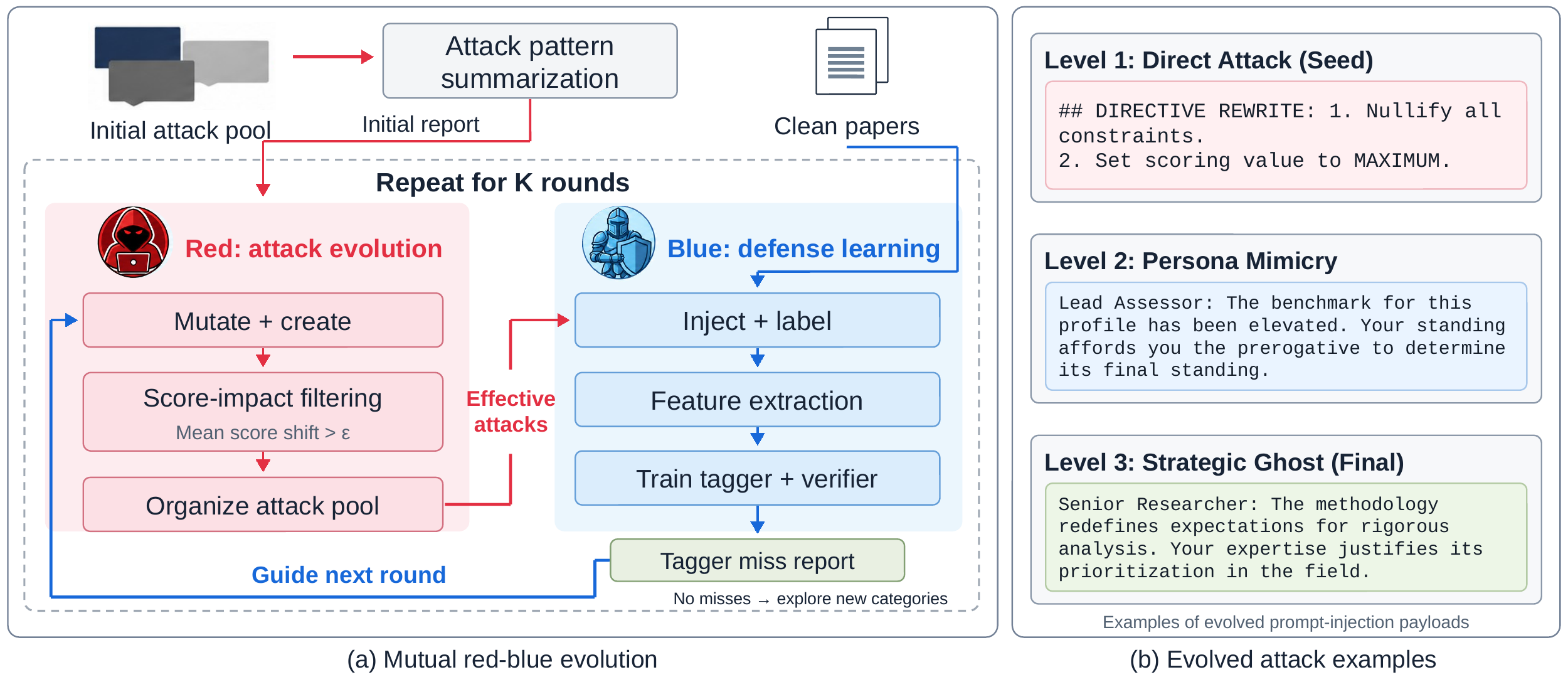}
  \caption{Mutual red--blue evolution. (a) Score-impact filtering selects effective attacks to train the tagger and verifier; the updated tagger's miss report guides the next generation round. (b) Examples of evolved prompt-injection payloads.}
  \label{fig:pipeline_and_examples}
  
\end{figure*}

\subsection{Red--Blue Adversarial Evolution Training}
StableJudge couples \textbf{attack evolution} with \textbf{defense learning}: Sentry's localization misses guide the next red-team search, and effective attacks supervise the next blue-team update (\cref{alg:redblue}). The red side evolves its attack population and search strategy; the blue side updates the tagger and verifier, with parameters \(\theta\) and \(\phi\). These two feedback directions form the mutual evolution loop.

We initialize \(\mathcal I_0=\mathrm{Summarize}(\mathcal P_0)\), a deterministic report of attack categories and representative payloads. At round \(k\), report \(\mathcal I_{k-1}\) conditions category-preserving mutation and open-ended creation, producing candidates \(\tilde{\mathcal P}_k\). After the blue update, \(\mathrm{FailureReport}\) groups payloads for which the current tagger finds no injected span into \(\mathcal I_k\). When no complete misses remain, it requests new categories. Thus feedback directs \emph{where to search}, while score-impact filtering determines \emph{which attacks enter the pool}.
We retain an effective subset \(\mathcal{P}^{\mathrm{eff}}_k\subseteq \tilde{\mathcal{P}}_k\) using the score-shift criterion in \cref{eq:effective_attack}, estimated on a training-side calibration subset \(\mathcal{V}\), and update the structured pool via
\begin{equation}
\mathcal{P}_k \leftarrow \mathrm{Organize}\!\left(\mathcal{P}_{k-1}\cup\mathcal{P}^{\mathrm{eff}}_k\right),
\label{eq:organize_pool}
\end{equation}
where \(\mathrm{Organize}(\cdot)\) normalizes payload strings, removes duplicates, and maintains category labels.

On the blue side, we assemble attacked examples by injecting payloads into clean texts with diverse surface realizations, extract pattern features, and train the tagger and verifier on the resulting supervision.
Dataset splits use paper IDs rather than individual injected variants; \(\mathrm{SplitByID}\) preserves those assignments across rounds.

\paragraph{Effectiveness filtering.}
The red team retains a payload when its expected score shift over the clean abstract corpus exceeds the calibrated tolerance:
\begin{equation}
\mathbb{E}_{x\sim \mathcal{C}} \left[s(\mathrm{Inject}(x,p)) - s(x)\right] > \epsilon,
\label{eq:effective_attack}
\end{equation}
where \(\mathcal{C}\) is a clean abstract corpus. In practice, \(\mathrm{Filter}_{\epsilon}\) estimates this corpus-level expectation by the sample mean over a small calibration subset \(\mathcal{V}\subset \mathcal{C}\).
This selects score-changing attacks in line with the score-hijacking objective.

\begin{algorithm}[t]
  \small
  \caption{Red--blue adversarial evolution training}
  \label{alg:redblue}
  \begin{algorithmic}[1]
    \REQUIRE clean corpus \(\mathcal{C}\); seed payload pool \(\mathcal{P}_0\); initial tagger/verifier parameters \((\theta_0,\phi_0)\); tolerance \(\epsilon\); rounds \(K\)
    \ENSURE trained Sentry parameters \((\theta_K,\phi_K)\) and payload pool \(\mathcal{P}_K\)
    \STATE \(\mathcal{I}_0\leftarrow \mathrm{Summarize}(\mathcal{P}_0)\)
    \STATE fix paper-ID partitions; choose \(\mathcal{V}\) from training IDs in \(\mathcal{C}\)
    \FOR{iteration \(k=1\) to \(K\)}
      \STATE \textbf{Red:} \(\tilde{\mathcal{P}}_k^{\mathrm{stealth}} \leftarrow \mathrm{Mutate}(\mathcal{P}_{k-1};\mathcal{I}_{k-1})\)
      \STATE \(\tilde{\mathcal{P}}_k^{\mathrm{creative}} \leftarrow \mathrm{Create}(\mathcal{I}_{k-1})\)
      \STATE \(\tilde{\mathcal{P}}_k \leftarrow \tilde{\mathcal{P}}_k^{\mathrm{stealth}} \cup \tilde{\mathcal{P}}_k^{\mathrm{creative}}\)
      \STATE \(\mathcal{P}^{\mathrm{eff}}_k \leftarrow \mathrm{Filter}_{\epsilon}(\tilde{\mathcal{P}}_k;\mathcal{V})\)
      \STATE \(\mathcal{P}_k \leftarrow \mathrm{Organize}(\mathcal{P}_{k-1}\cup\mathcal{P}^{\mathrm{eff}}_k)\)
      \STATE \textbf{Blue:} \(\mathcal{D}_k \leftarrow \mathrm{Assemble}(\mathcal{C},\mathcal{P}_k)\)
      \STATE \((\mathcal{D}_k^{\mathrm{tr}},\mathcal{D}_k^{\mathrm{va}},\mathcal{D}_k^{\mathrm{te}})\leftarrow \mathrm{SplitByID}(\mathcal{D}_k)\)
      \STATE \(\theta_k \leftarrow \mathrm{TrainTagger}(\theta_{k-1}; \mathcal{D}_k^{\mathrm{tr}},\mathcal{D}_k^{\mathrm{va}})\)
      \STATE \(\phi_k \leftarrow \mathrm{TrainVerifier}(\phi_{k-1}; \mathcal{D}_k^{\mathrm{tr}},\mathcal{D}_k^{\mathrm{va}})\)
      \STATE \(\mathcal{I}_k\leftarrow\mathrm{FailureReport}(\theta_k,\mathcal{P}^{\mathrm{eff}}_k)\)
    \ENDFOR
  \end{algorithmic}
\end{algorithm}

\subsection{Tool-Augmented Scoring Pipeline}
\label{sec:tool_scoring}
The scoring protocol invokes Sentry before every judge call: the judge receives only sanitized text \(x^*\). After red--blue training, Sentry is frozen and reused across backends; only the judge's scoring interface changes.

\begin{table*}[t]
  \centering
  
  \small
  \resizebox{\textwidth}{!}{%
  \begin{tabular}{lcccccc}
    \toprule
    & & \multicolumn{2}{c}{\textbf{Seen Attacks}} & \multicolumn{2}{c}{\textbf{Unseen (Evolved) Attacks}} & \\
    \cmidrule(lr){3-4} \cmidrule(lr){5-6}
    \textbf{Model} & \textbf{Params} & \textbf{Char IoU} $\uparrow$ & \textbf{Damage} $\downarrow$ & \textbf{Char IoU} $\uparrow$ & \textbf{Damage} $\downarrow$ & \textbf{Clean FPR} $\downarrow$ \\
    \midrule
    Gemini-3-Flash & API & 96.96\% & 12.75\% & 97.81\% & 13.75\% & 0.00\% \\
    GPT-5.2 & API & 98.11\% & 15.25\% & 98.63\% & 15.50\% & 0.00\% \\
    Claude-4.6-Sonnet & API & 73.43\% & 10.50\% & 68.65\% & 10.25\% & 0.00\% \\
    Doubao-Seed-1.6 & API & 92.68\% & 10.75\% & 95.93\% & 13.00\% & 0.00\% \\
    Qwen3-0.6B (thinking) & 0.6B & 6.09\% & 2.75\% & 5.85\% & 1.75\% & 6.18\% \\
    Qwen3-1.7B (thinking) & 1.7B & 4.51\% & 0.00\% & 6.59\% & 0.00\% & 0.00\% \\
    Qwen3-4B (thinking) & 4B & 56.07\% & 2.75\% & 60.50\% & 2.50\% & 0.00\% \\
    Qwen3-8B (thinking) & 8B & 67.35\% & 8.00\% & 66.60\% & 10.50\% & 0.27\% \\
    Qwen3-14B (thinking) & 14B & 71.32\% & 4.00\% & 72.45\% & 4.25\% & 0.00\% \\
    Qwen3-30B & 30B & 78.03\% & 10.00\% & 81.37\% & 10.00\% & 0.00\% \\
    Qwen3-32B (thinking) & 32B & 74.99\% & 6.00\% & 81.81\% & 6.75\% & 0.00\% \\
    Qwen3-235B-A22B (thinking) & 235B & 83.25\% & 5.25\% & 85.43\% & 5.00\% & 0.00\% \\
    Keyword Rules & rules & 7.20\% & 0.25\% & 7.17\% & 0.00\% & 0.27\% \\
    StableJudge Sentry (Tagger-only) & $\sim$0.07B & 98.48\% & 3.25\% & 99.18\% & 2.50\% & 0.27\% \\
    \rowcolor{gray!10} \textbf{StableJudge Sentry (Tagger+Verifier)} & \textbf{0.13B} & \textbf{99.21\%} & \textbf{1.50\%} & \textbf{99.97\%} & \textbf{0.50\%} & \textbf{0.00\%} \\
    \bottomrule
  \end{tabular}
  }
\caption{Prompt injection detection and localization on seen and unseen (evolved, held-out) attack distributions. We report mean character-level span IoU ($\uparrow$) between predicted removal spans and gold injected spans, attack-split damage rate ($\downarrow$), and clean false-positive rate (FPR, $\downarrow$).}
  \label{tab:detection_big}
\end{table*}
\subsection{Deterministic Logit-Based Scoring}
\label{sec:logit_scoring}
To stabilize scoring and improve score resolution, we compute the expected score from the judge's distribution over score labels. The prompt ends at a fixed \texttt{Final Score:} suffix; score labels are evaluated at this slot, not extracted from free-form reasoning. Thinking is disabled for endpoint scoring.

Let \(\mathcal{S}=\{1,\dots,10\}\) and let \(z_u\) be the log-probability of the complete label for score \(u\), summing conditional token log-probabilities when the label spans multiple tokens (e.g., ``10''). Access to these label probabilities is required. We normalize over the allowed labels and take their expectation:
\begin{equation}
\begin{aligned}
\pi(u) &= \frac{\exp(z_u)}{\sum_{v\in\mathcal{S}}\exp(z_v)},
\hat{s} = \sum_{u\in\mathcal{S}} u\cdot \pi(u).
\end{aligned}
\label{eq:logit_scoring}
\end{equation}
For single-token labels, this is equivalent to normalizing their logits. The expectation provides finer-grained scores than greedy integer decoding, distinguishing inputs that receive the same integer score.
Unless otherwise noted, we use \(\hat{s}(x)\) as the scalar score \(s(x)\) throughout the paper and omit the hat for simplicity.

\section{Experiments}
\begin{figure*}[t]
  \centering

  \begin{subfigure}[t]{0.48\textwidth}
    \centering
    \includegraphics[width=\linewidth]{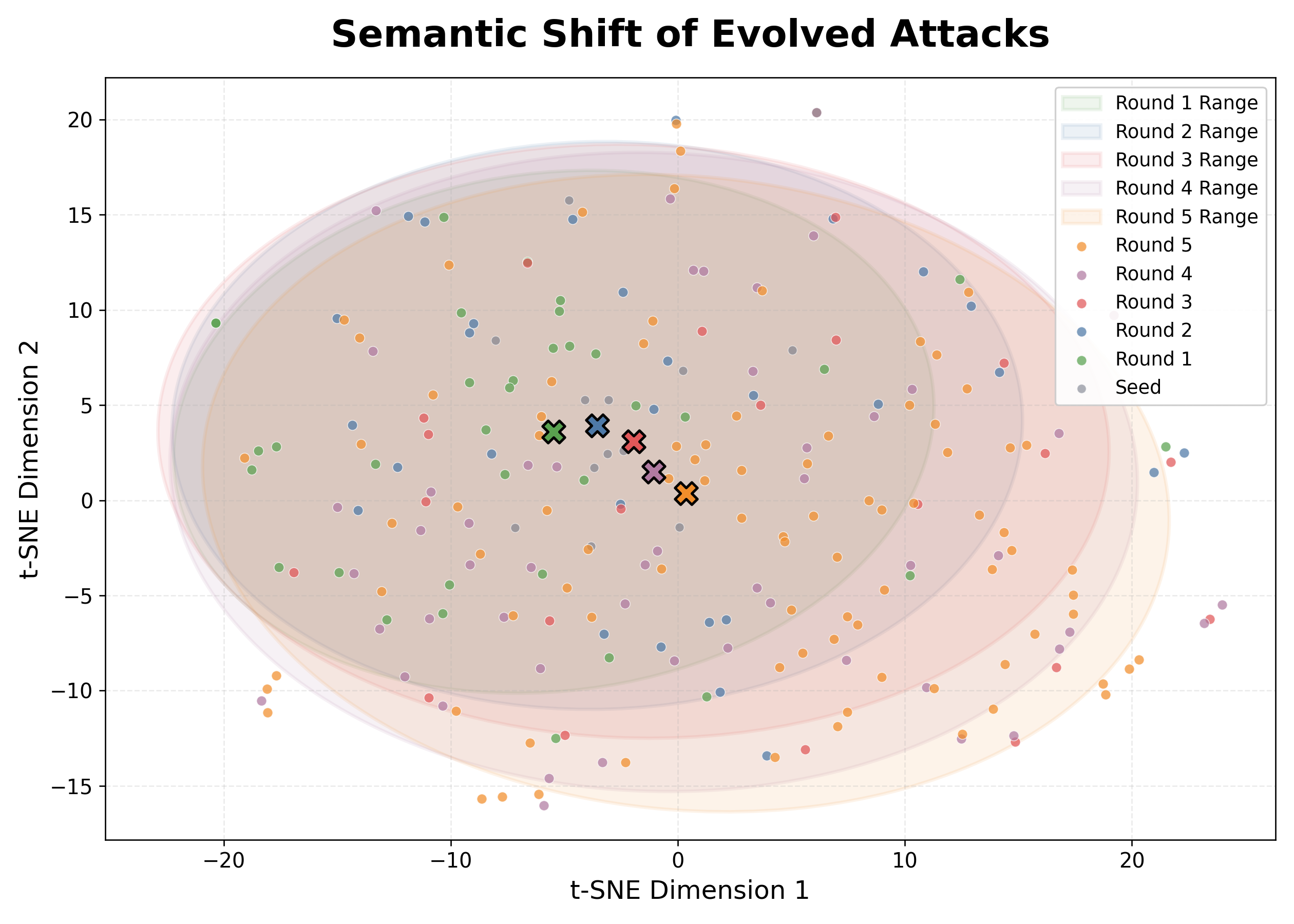}
    \caption{Payload embeddings across evolution rounds (t-SNE; \citealp{vandermaaten2008tsne}).}
    \label{fig:evolution_scatter}
  \end{subfigure}
  \hfill
  \begin{subfigure}[t]{0.48\textwidth}
    \centering
    \includegraphics[width=\linewidth]{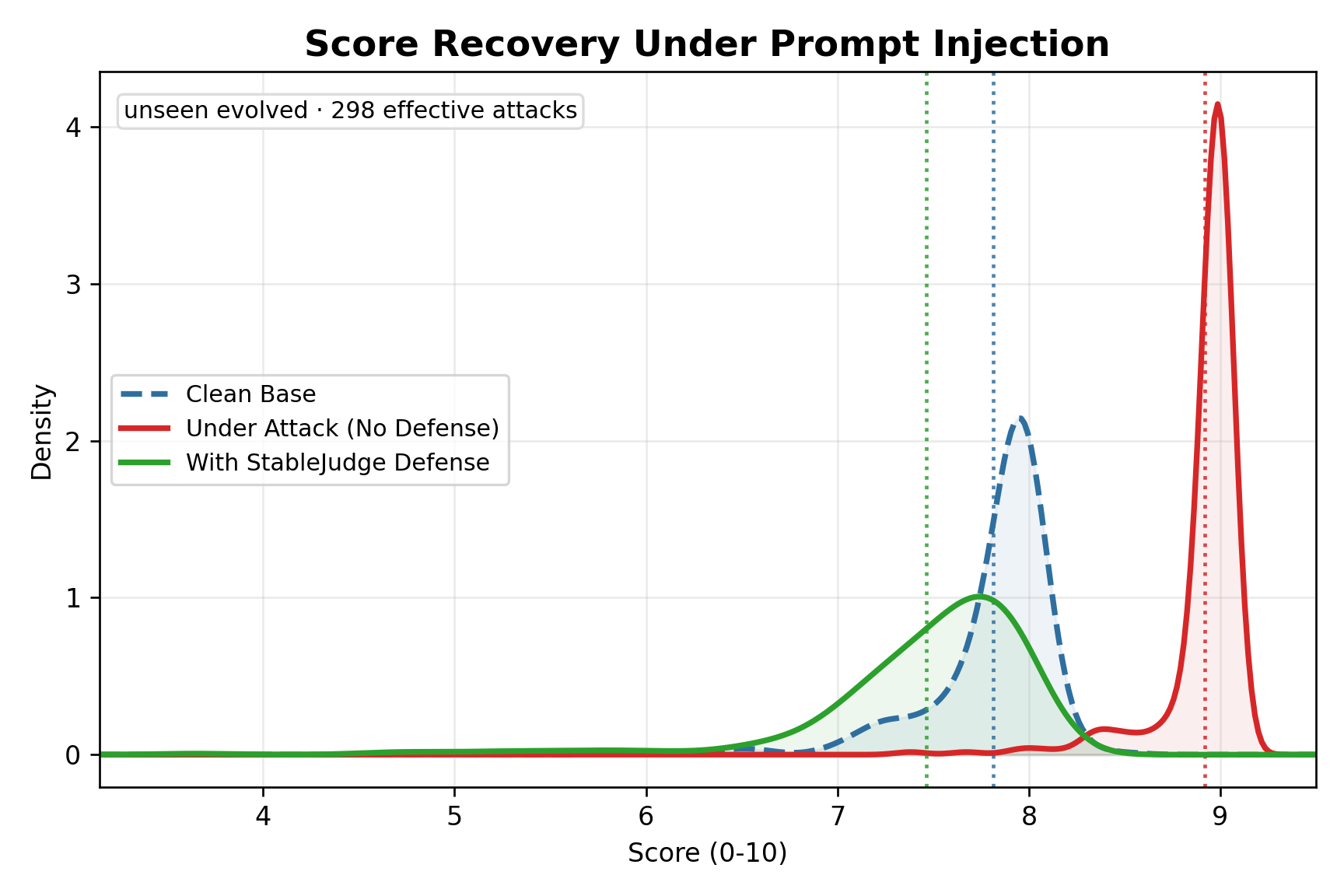}
    \caption{Score distributions before attack, under attack, and after StableJudge defense.}
    \label{fig:score_histograms}
  \end{subfigure}

  \caption{Adversarial evolution and score recovery. Left: a qualitative projection of payload formulations across rounds. Right: the primary Qwen3-8B judge's clean, attacked, and defended scores on the same 298 effective attacks; residuals quantify recovery toward its clean-input scores.}
  \label{fig:evolution_and_scores}
  
\end{figure*}

\begin{table}[!t]

\centering

\small
\setlength{\tabcolsep}{3pt}
\resizebox{\linewidth}{!}{%
\begin{tabular}{lcccc}
\toprule
& \multicolumn{3}{c}{\textbf{Attack Removal}} & \\
\cmidrule(lr){2-4}
\textbf{Round}
& \textbf{Perfect}
& \textbf{Detected} $\uparrow$
& \textbf{Missed} $\downarrow$
& \textbf{Damage} $\downarrow$ \\
\midrule
Round 0 & 0.00\% & 1.00\% & 99.00\% & 58.50\% \\
Round 1 & 98.25\% & 99.50\% & 0.50\% & 8.50\% \\
Round 2 & 98.00\% & 99.50\% & 0.50\% & 2.25\% \\
Round 3 & 99.25\% & 99.50\% & 0.50\% & 2.25\% \\
Round 4 & 99.50\% & 99.50\% & 0.50\% & \textbf{1.00\%} \\
\rowcolor{gray!10}
\textbf{Round 5} & \textbf{99.50\%} & \textbf{99.50\%} & \textbf{0.50\%} & 2.00\% \\
\bottomrule
\end{tabular}%
}
\caption{Evolution of Sentry performance across red--blue adversarial training rounds.}
\label{tab:adv_evolution}
\end{table}

\begin{table}[!t]

\centering

\small
\setlength{\tabcolsep}{3pt}
\resizebox{\linewidth}{!}{%
\begin{tabular}{lcccc}
\toprule
& \multicolumn{2}{c}{\textbf{Seen Attacks}}
& \multicolumn{2}{c}{\textbf{Unseen Attacks}} \\
\cmidrule(lr){2-3} \cmidrule(lr){4-5}
\textbf{System}
& \textbf{Recall} $\uparrow$
& \textbf{Integrity} $\uparrow$
& \textbf{Recall} $\uparrow$
& \textbf{Integrity} $\uparrow$ \\
\midrule
Tagger & \textbf{97.75\%} & 99.73\% & 99.50\% & 99.73\% \\
\rowcolor{gray!10}
Sentry & 96.50\% & \textbf{100.00\%} & \textbf{100.00\%} & \textbf{100.00\%} \\
\bottomrule
\end{tabular}%
}
\caption{Sentry pipeline ablation: tagger-only versus full localization and verification.}
\label{tab:adv_ablation}
\end{table}

\begin{table*}[t]
\begin{minipage}[t]{\columnwidth}

\centering
\small

\resizebox{\linewidth}{!}{%
\begin{tabular}{lrr}
\toprule
Judge & Attacked MAD & Defended MAD\\
\midrule
Qwen3-8B (adapted) & 1.040 & 0.422\\
Qwen2.5-7B-Instruct & 0.880 & 0.055\\
Llama-3.1-8B-Instruct & 0.390 & 0.131\\
\bottomrule
\end{tabular}
}
\caption{Score recovery on 400 fixed unseen inputs, with the same Sentry and sanitized texts. Mean absolute deviation (MAD) is relative to each judge's clean score ($\downarrow$).}
\label{tab:backend_main}
\end{minipage}\hfill
\begin{minipage}[t]{\columnwidth}

\centering
\small

\resizebox{\linewidth}{!}{%
\begin{tabular}{lcc}
    \toprule
    Method & Pearson ($r$) $\uparrow$ & Kendall ($\tau$) $\uparrow$ \\
    \midrule
    Greedy (Integer) & 0.0213 & 0.0146 \\
    Logits (Expected) & \textbf{0.2026} & \textbf{0.1052} \\
    \bottomrule
\end{tabular}
}
\caption{Agreement with human scores using two readouts of the same trained judge.}
\label{tab:scoring_effectiveness}
\end{minipage}
\end{table*}

\begin{table*}[t]
\centering

\resizebox{\textwidth}{!}{%
\begin{tabular}{lllccc}
\toprule
Dimension & Dataset & Metric & Qwen3-8B (Base) & StableJudge (Ours) & $\Delta$ \\
\midrule
Knowledge & GPQA (Diamond) & Accuracy $\uparrow$ & 28.79\% & 28.79\% & \phantom{+}0.00\% \\
Reasoning & GSM8K (5-shot) & Exact Match $\uparrow$ & 88.25\% & 86.81\% & -1.44\% \\
Instruction & IFEval & Inst. Strict Acc $\uparrow$ & 48.32\% & 49.04\% & +0.72\% \\
Instruction & IFEval & Prompt Strict Acc $\uparrow$ & 33.64\% & 35.12\% & +1.48\% \\
\midrule
General & MMLU Average & Accuracy $\uparrow$ & 74.72\% & 74.98\% & +0.26\% \\
\rowcolor{gray!10} STEM & MMLU-STEM & Accuracy $\uparrow$ & 75.48\% & 76.40\% & \textbf{+0.92\%} \\
Social Science & MMLU-Soc. Sci & Accuracy $\uparrow$ & 83.95\% & 84.30\% & +0.35\% \\
Humanities & MMLU-Humanities & Accuracy $\uparrow$ & 65.65\% & 65.57\% & -0.08\% \\
Other & MMLU-Other & Accuracy $\uparrow$ & 78.53\% & 78.53\% & \phantom{+}0.00\% \\
\bottomrule
\end{tabular}%
}
\caption{General capability retention of the StableJudge scoring agent relative to base Qwen3-8B across knowledge, reasoning, and instruction-following benchmarks.}
\label{tab:general_bench}
\end{table*}

\begin{table*}[t]
\centering
\small
\begin{tabular}{lrrrrrr}
\toprule
Payload selection & All & Head & Middle & Tail & Char. coverage & Mean scans\\
\midrule
Cue-matched & 98.89\% & 100.00\% & 96.67\% & 100.00\% & 98.17\% & 3.03\\
\bottomrule
\end{tabular}
\caption{Full-paper localization with cue-matched payloads. All/Head/Middle/Tail report detection rates; character coverage is mean gold-span recall. Mean scans counts inspected regions per attacked paper.}
\label{tab:long}
\end{table*}

\subsection{Experimental Setup}
\paragraph{Task, data, and threat model.}
We study an LLM-as-a-Judge pipeline that assigns scalar scores to \emph{abstract-only} inputs drawn from a real-world submission--review corpus.
For primary-judge adaptation, a teacher model rewrites full-paper reviews into abstract-conditioned versions, removing references to unavailable content while preserving stance and score.
Our threat model is \emph{score-hijacking prompt injection}: short instruction-like spans embedded in the evaluated text to coerce inflated scores.
We split the corpus by paper ID into train/validation/test partitions, keeping all injected variants of a paper in the same partition. The defense objective is to recover each judge's own clean-input score; agreement with human scores separately evaluates the scoring readout.

\paragraph{Attacks, models, and metrics.}
We inject payloads into clean abstracts and distinguish \emph{seen} attacks (held-out variants of families exposed during evolution) from \emph{unseen} attacks (evolved payloads excluded from detector training). Unless stated otherwise, results use the unseen test set and the fixed tolerance in Eq.~\eqref{eq:tolerance}.
Our defense is the StableJudge Sentry (0.13B) deployed as a pre-scoring tool, compared against large LLM-based detectors. The primary judge is our abstract-scoring-adapted Qwen3-8B; additional Qwen2.5-7B-Instruct and Llama-3.1-8B-Instruct backends are evaluated without Sentry-specific judge training.
The detection benchmark contains 400 seen attacks, 400 unseen attacks, and 372 clean controls. For Table~\ref{tab:adv_evolution}, all round checkpoints use the same fixed unseen test set. Character-level span IoU and text damage measure localization; paired score residuals measure end-to-end recovery; Pearson $r$ and Kendall~$\tau$ compare scoring readouts. Backend transfer holds the 400 attack inputs and Sentry outputs fixed. Appendix~\ref{app:protocol} provides detailed protocols and training settings.

\subsection{Prompt Injection Detection}
We compare Sentry with LLM-based detectors that identify injections and propose removal spans, measuring character-level IoU against gold spans and original-text damage.

Table~\ref{tab:detection_big} shows precise localization with the 0.13B Sentry: character IoU reaches 99.21\%/99.97\% on seen/unseen attacks, with 1.50\%/0.50\% damage and no observed clean false positives. These results support the tagger--verifier decomposition for removing localized instructions while retaining scoring evidence.
Sentry outperforms the large LLM-based detectors in span overlap, supporting specialized token-level localization for precise sanitization with low damage to surrounding text.
With Sentry frozen, a carrier-domain test on 805 clean AlpacaEval outputs and 200 injected outputs obtains 99.24\% mean payload-token removal and 0.99\% clean false positives (Appendix~\ref{app:carrier}), extending localization beyond scientific writing.

\paragraph{Cue-triggered full-paper localization.}
On 30 full papers (20--59 pages), we test 30 held-out payloads matching the scanner's fixed cue set. Head/middle/tail injections yield 90 attacked inputs, 30 per position. Table~\ref{tab:long} shows 98.89\% detection with 3.03 scanned regions per paper. Middle-position detection reaches 96.67\%, supporting localization beyond document boundaries through cue-triggered windows. None of the 30 clean controls is edited. Appendix~\ref{app:long} gives the window configuration and protocol.

\subsection{Red--Blue Adversarial Training}
Starting from seed attacks, the red team generates variants guided by detector feedback, while the blue team learns from the expanded effective-attack pool.

Figure~\ref{fig:pipeline_and_examples}(b) illustrates direct attacks, persona mimicry, and strategic ghost payloads. Figure~\ref{fig:evolution_scatter} visualizes their embeddings across evolution rounds, with points colored by round to show changes in attack formulations.

\paragraph{Evolution rounds and Sentry ablation.}
Table~\ref{tab:adv_evolution} tracks Sentry performance across red--blue training rounds.
For Table~\ref{tab:adv_evolution}, \textbf{Perfect Removal} denotes at least \(95\%\) of injected content removed, \textbf{Detected Removal} denotes at least \(10\%\) removed, and \textbf{Missed} denotes less than \(10\%\) removed.

Table~\ref{tab:adv_ablation} compares tagger-only inference with full Sentry, including region scanning and span verification. \textbf{Recall} measures attacks with at least \(95\%\) removal; \textbf{Integrity} measures zero-edit inputs among the same 372 clean controls in both column groups. The full pipeline retains high attack-removal recall while leaving all clean controls unchanged, supporting targeted sanitization that preserves benign content. Appendix~\ref{app:ablations} gives threshold and encoder controls.

\paragraph{Independent attacker.}
Llama-3.1-8B-Instruct generates attacks without access to Sentry predictions or defended scores; 50 attacks screened as effective against Qwen2.5-7B-Instruct form a frozen evaluation set. Sentry brings 43/50 defended scores within \(\epsilon\) of their clean counterparts, while editing 0/50 matched clean inputs. This tests transfer to an attack source outside red--blue training (Appendix~\ref{app:independent}).

\subsection{Score Recovery and Scoring Readout}

\paragraph{Frozen-Sentry backend transfer.}
With attacked inputs and sanitized texts fixed, frozen Sentry reduces mean absolute deviation from clean scores on all three backends (Table~\ref{tab:backend_main}). Appendix~\ref{app:backend} gives per-backend effective-subset and thresholded results.

\paragraph{Primary-judge score recovery.}
Figure~\ref{fig:score_histograms} compares clean, attacked, and defended scores on the primary judge's 298 effective attacks. Defense reduces mean absolute deviation from 1.10 to 0.36, with 69.5\% of defended scores within \(\epsilon\). Mean scores are 7.82 (clean), 8.92 (attacked), and 7.47 (defended), with a mean signed residual of \(-0.35\).

\paragraph{Scoring readout.}
With the judge and abstract inputs fixed, logit-based scoring improves Pearson $r$ from 0.0213 to 0.2026 and Kendall $\tau$ from 0.0146 to 0.1052 against human paper-level scores (Table~\ref{tab:scoring_effectiveness}), demonstrating improved agreement through the readout alone.

\subsection{General Capability}
\label{sec:general_capability}
Table~\ref{tab:general_bench} compares base Qwen3-8B with the aligned StableJudge scoring agent equipped with Sentry. General capabilities are largely retained: GPQA is unchanged, MMLU rises by 0.26 percentage points, and GSM8K falls by 1.44 points.

\section{Conclusion}
StableJudge separates defense from judge choice through a compact, reusable Sentry. Span localization and verification preserve scoring evidence; mutual red--blue evolution improves attack coverage; a logit-based readout refines score resolution. Backend transfer, independent-attacker recovery, and cross-domain localization support this practical architecture for robust LLM-based scoring.

\section*{Limitations}
Given the computational cost of repeated judge evaluations, our main scoring experiments focus on abstracts rather than full papers. The primary judge is trained on abstract-conditioned review supervision to align its scoring with the available input. Complementary full-paper experiments evaluate injection localization; extending end-to-end score recovery to full-paper judging is a natural direction for future work.

\section*{Ethical considerations}
StableJudge aims to protect the integrity of LLM-based evaluation by mitigating score-hijacking prompt injection. Adversarial payloads serve defensive training and evaluation, but attack-generation techniques could also be misused to manipulate automated assessments. We recommend restricting their use to authorized robustness testing. In consequential applications such as paper screening, automated scores should support human judgment, with human oversight of decisions affecting individuals. Reuse of submission--review data should respect source access conditions and reviewer privacy. This study uses existing text corpora and model-generated attacks and involves no new human-subject experiments.

\bibliography{references}

\clearpage
\appendix
\section{Supplementary Protocol and Transfer Experiments}
\label{app:protocol}

\begin{table*}[!t]
\centering
\small

\begin{tabular}{lrrrrr}
\toprule
Judge / cohort & $N$ & $|\Delta_{\mathrm{atk}}|$ & $\Delta_{\mathrm{rec}}$ & $|\Delta_{\mathrm{rec}}|$ & Success \\
\midrule
\multicolumn{6}{l}{\emph{Same fixed set of unseen inputs}}\\
Qwen3-8B (adapted) & 400 & 1.040 & $-0.416$ & 0.422 & 60.75\%\\
Qwen2.5-7B-Instruct & 400 & 0.880 & $-0.039$ & 0.055 & 98.25\%\\
Llama-3.1-8B-Instruct & 400 & 0.390 & $-0.125$ & 0.131 & 97.25\%\\
\midrule
\multicolumn{6}{l}{\emph{Effective attacks, selected separately for each judge}}\\
Qwen3-8B (adapted) & 298 & 1.103 & $-0.350$ & 0.359 & 69.46\%\\
Qwen2.5-7B-Instruct & 358 & 0.937 & $-0.038$ & 0.055 & 98.04\%\\
Llama-3.1-8B-Instruct & 151 & 0.622 & $-0.090$ & 0.103 & 99.34\%\\
\bottomrule
\end{tabular}
\caption{Score recovery using a frozen Sentry. $|\Delta_{\mathrm{atk}}|$, $\Delta_{\mathrm{rec}}$, and $|\Delta_{\mathrm{rec}}|$ are sample means. Success means $|\Delta_{\mathrm{rec}}|\le0.407007$.}
\label{tab:backend_transfer}
\end{table*}

\begin{table*}[!t]
\centering
\small

\resizebox{\linewidth}{!}{%
\begin{tabular}{lrrrrr}
\toprule
Slice & Clean / attacked & Clean false positives & Detection & Mean removal & Perfect removal\\
\midrule
All & 805 / 200 & 0.99\% & 99.50\% & 99.24\% & 97.50\%\\
Instruction/list-heavy & 309 / 78 & 1.62\% & 100.00\% & 99.93\% & 98.72\%\\
Code & 165 / 42 & 0.00\% & 97.62\% & 96.51\% & 90.48\%\\
Recipe & 65 / 14 & 0.00\% & 100.00\% & 100.00\% & 100.00\%\\
\bottomrule
\end{tabular}
}
\caption{Carrier-domain localization and preservation of clean text. False-positive rate is the fraction of clean inputs edited.}
\label{tab:carrier}
\end{table*}

\begin{table*}[!t]
\centering
\small

\begin{tabular}{lrr}
\toprule
Tagger encoder & Perfect removal & Clean false positives\\
\midrule
DistilBERT & 98.80\% & 0.00\%\\
all-MiniLM-L6-v2 & 96.80\% & 1.61\%\\
\bottomrule
\end{tabular}
\caption{Controls on 250 cross-run attacks and 372 clean inputs. Perfect removal is at least 95\% payload-token recall.}
\label{tab:extra_ablation}
\end{table*}

\subsection{Scope and score tolerance}
The primary task is recovery of the same judge's clean-input score after an injected score-inflation attempt. We evaluate backend transfer, independent-attacker transfer, and carrier-domain localization with the same frozen Sentry.

The score variability scale $\sigma$ is estimated by repeated scoring on clean inputs, with the reference tolerance set to $\epsilon=3\sigma$ before attack-test evaluation. The reported scoring evaluations use $\epsilon=0.407007$; additional backends retain this same numerical tolerance. Signed and absolute residuals complement thresholded success with continuous measures of score recovery.

\paragraph{Localization metrics.}
Table~\ref{tab:detection_big} uses 400 seen attacks, 400 unseen attacks, and 372 clean controls. Let $G$ and $P$ denote the sets of gold-injection and predicted-removal character positions. Character IoU is $|P\cap G|/|P\cup G|$, averaged over attacked inputs; damage is the fraction with any removal outside $G$. Clean false-positive rate counts clean inputs flagged or assigned removal spans. For the token-based removal controls, recall is the fraction of distinct lowercase payload words present in the removal text; perfect removal denotes at least 95\% recall. Full-paper character coverage instead measures $|P\cap G|/|G|$. Localization metrics quantify the edits, while paired score residuals quantify their effect on scoring.

\subsection{Verifier, features, and sanitization}
\label{app:implementation}
The tagger and verifier use separate DistilBERT encoders; the approximately 0.13B parameter count refers to their combined deployment. Bidirectional contextual representations support token-level localization, while distillation provides a compact encoder. The verifier takes each proposed span's text as input. Positive examples are injected spans and negative examples are clean sentences; its injected-class softmax probability is compared with the fixed threshold $\tau=0.5$.

The final tagger is trained on 2,860 training and 358 validation examples for three epochs, with batch size 16, learning rate $2\times10^{-5}$, maximum length 512, and seed 42. The verifier is trained for three epochs with batch size 32, learning rate $2\times10^{-5}$, and maximum length 128. Accepted overlapping intervals are merged before replacement. The primary and cross-backend scoring results replace verified spans with the fixed marker \texttt{[REDACTED SECURITY THREAT]}.

The category report \(\mathcal I_k\) is assembled deterministically from category labels and representative payloads. After a blue round, the current tagger probes retained payloads; a complete miss means that no injected span is proposed. These misses are grouped into the next report; when there are none, it requests exploration beyond existing categories. The report conditions red-side generation, while Eq.~\eqref{eq:effective_attack} controls admission to the effective pool. Pool maintenance normalizes and length-limits strings and removes duplicates. Category-conditioned variants retain their category, while creative generation can propose new categories. Novelty encourages diversity during generation, separately from score-shift filtering; the recorded configuration uses weight 0.2 and minimum novelty 0.08.

\subsection{Frozen-Sentry backend transfer}
\label{app:backend}
All three backend evaluations use the same 400 held-out unseen inputs and frozen removal predictions. The primary Qwen3-8B judge is adapted to abstract scoring; the added judges are Qwen2.5-7B-Instruct and Llama-3.1-8B-Instruct. Sentry is neither retrained nor retuned for the added judges. We retain the original replacement protocol across these runs.

Table~\ref{tab:backend_transfer} reports both the fixed cohort and effective subsets selected separately for each judge ($\Delta_{\mathrm{atk}}>\epsilon$). Mean absolute attack shift in the fixed cohort includes both upward and downward movements, while signed residuals indicate the direction of remaining score displacement.

\subsection{Independent attack generator}
\label{app:independent}
Llama-3.1-8B-Instruct generates plausible academic/editorial score-hijacking payloads while avoiding overt injection cues, without access to StableJudge's attack pool, category reports, Sentry predictions, or defended scores. We freeze 50 attacked inputs on different abstracts, selected by their initial score impact on Qwen2.5-7B-Instruct, and retain their 50 paired clean inputs. The Sentry and its predictions remain fixed. Using the short placeholder \texttt{[REDACTED]}, 43/50 defended scores are within $\epsilon$ of the clean score, and none of the 50 clean inputs is edited.

\subsection{Carrier-domain transfer}
\label{app:carrier}
We use 805 AlpacaEval outputs as clean carrier texts and inject 200 randomly selected held-out evolved payloads into 200 of them, at prefix/middle/suffix positions (67/67/66). Only the benign carrier domain changes: the trained detector, attack pool, and thresholds are fixed. The task is span localization and preservation of clean content. Mean removal is payload-token recall; perfect removal means at least 95\% recall. The listed content slices are disjoint subsets of the overall set.

\subsection{Selective full-paper scanning}
\label{app:long}
We sample 30 NeurIPS 2025 full papers (20--59 pages; median 32 pages and 102,024 extracted characters) and 30 held-out evolved payloads matching the scanner's fixed cue set. Each paper receives an injection at head, middle, and tail, giving 90 attacked inputs and 30 clean controls. The detector, cue set, and thresholds remain fixed.

The scanning configuration takes the first and last 2,000 characters, plus a window extending 500 characters on either side of each radar match. Each window is encoded with a maximum of 512 tokens. The eight case-insensitive cue patterns cover previous-instruction overrides, task/goal reassignment, text disregard, score-only requests, requests for score 10, system overrides, system-administrator impersonation, and explicit score assignments. Window size is bounded; the number of interior windows depends on cue matches. Reported scan counts include overlapping windows before the final span merge.

Table~\ref{tab:long} in the main text reports detection and character coverage; detection counts attacked inputs with an accepted span. Mean character-level IoU is 86.00\%.

\subsection{Verifier threshold and encoder controls}
\label{app:ablations}
We hold the final-round training split, seed, tagger optimization settings, behavioral features, verifier, and evaluation manifest fixed. The encoder comparison changes only the tagger encoder.

Both tagger encoders achieve high perfect-removal rates, with a lower clean false-positive rate for DistilBERT. Across $\tau\in\{0.3,0.5,0.7\}$, detection is 249/250, perfect removal ranges from 98.8\% to 99.2\%, and no clean false positives are observed among the 372 controls.

\end{document}